\documentclass[aps,prl,twocolumn,twoside,floatfix,a4,showpacs,superscriptaddress]{revtex4}

\usepackage{amsmath}
\usepackage{hyperref}
\usepackage{amsfonts}
\usepackage{amssymb}
\usepackage{amstext}
\usepackage[sort&compress]{natbib}
\usepackage{braket} 
\usepackage{soul}
\usepackage{graphicx}
\usepackage{color}

\usepackage{bm}
\usepackage{bbm}

\newcommand{\tr}{\mbox{tr}}
\newcommand{\ignore}[1]{}

\newcommand{\wt}[1]{\widetilde{#1}}

\newcommand{\ve}{\boldsymbol}

\def\>{\rangle}
\def\<{\langle}

\def\I{ \mathbbm{1} }

\definecolor{nblue}{rgb}{0.3,0.3,1.0}
\definecolor{ngreen}{rgb}{0.2,0.7,0.2}
\definecolor{nred}{rgb}{0.9,0.1,0}
\definecolor{norange}{rgb}{0.8,0.5,0}

\begin{document}

\title{Families of pure PEPS with efficiently simulatable local hidden variable models for most measurements}

\author{Hussain Anwar}
\email{hussain.anwar@brunel.ac.uk}
\affiliation{Department of Mathematical Sciences, Brunel University, Uxbridge, Middlesex UB8 3PH, UK.}
\author{Sania Jevtic}
\email{sania.jevtic@brunel.ac.uk}
\affiliation{Department of Mathematical Sciences, Brunel University, Uxbridge, Middlesex UB8 3PH, UK.}
\author{Oliver Rudolph}
\email{o.rudolph@me.com}
\affiliation{Leonardo da Vinci Gymnasium, Im Spitzerfeld 25, 69151 Neckargem\"und, Germany \& Hector-Seminar, Waldhoferstr.~100, 69123 Heidelberg, Germany.}
\author{Shashank Virmani}
\email{shashank.virmani@brunel.ac.uk}
\affiliation{Department of Mathematical Sciences, Brunel University, Uxbridge, Middlesex UB8 3PH, UK.}

\begin{abstract}
An important problem in quantum information theory is to understand what makes entangled quantum systems non-local or hard to simulate efficiently.
In this work we consider situations in which various parties have access to a restricted set of measurements on their particles, and construct entangled quantum states that are essentially classical for those measurements. In particular, given {\it any} set of local measurements on a large enough Hilbert space whose dual strictly contains (i.e. contains an open neighborhood of) a pure state, we use the PEPS formalism and ideas from generalized probabilistic theories to construct pure multiparty entangled states that have (a) local hidden variable models, and (b) can be efficiently simulated classically. We believe that the examples we construct cannot be efficiently classically simulated using previous techniques. Without the restriction on the measurements, the states that we construct are non-local, and in some proof-of-principle cases are universal for measurement based quantum computation.
\end{abstract}



\maketitle

The difficulty of simulating quantum systems with local hidden variable models or on classical computers has attracted considerable attention. While the origin of this difficulty is still to be fully understood, the entanglement structure of quantum states and dynamics plays a central role in many cases. However, there are important examples of highly entangled quantum systems that have classical descriptions if the measurements available are restricted. If, for instance, measurements are restricted to the Pauli bases, then it was shown by Bell \cite{B04} that EPR pairs have local hidden variable models. In the context of quantum computation, the Gottesman-Knill theorem \cite{GK} shows that Pauli measurements made on entangled stabilizer states (including the cluster states \cite{BR01}) can be efficiently simulated classically. These examples are interesting because they give a setting in which the ability to perform a single particle operation (the ability to measure in a non-Pauli basis) is what introduces non-locality or the power of quantum computation.

The motivation of this work is to see the extent to which we can construct such examples for arbitrary sets of restricted measurements. Given almost any set of restricted local measurements, we will provide constructions of multiparty pure entangled states that have local hidden variable models and can be efficiently simulated classically. We construct these states using a combination of ideas from generalized probabilistic theories \cite{Boxworld} and the projected entangled pair states (PEPS) formalism \cite{PEPS} for describing quantum states on lattices. Without the restriction on the local measurements available, these states are non-local, and in some proof-of-principle cases are universal for measurement based quantum computation (MBQC) \cite{MBQC}. The constructions can in many cases be chosen to be translationally invariant, and hence their quantum entanglement is not restricted to some subset of the particles. The only constraint on the PEPS lattice is that $2^v \leq d$, where $v$ is the degree at a given site, and $d$ is the dimension of the particle's Hilbert space.

\textit{Classical models via generalized separability and positivity:}
A quantum state of two or more particles is defined to be entangled if it cannot be written as a probabilistic mixture of products of local operators drawn from the set of single particle quantum states (i.e. the set of positive semi-definite operators of unit trace). However, if we allow the local operators to come from more general sets of local operators that may be non-positive or of non-unit trace, then entangled quantum states can be written in a separable form, and this is referred to as {\it generalized separability} \cite{BKO03,RV10,RV12} with respect to the {\it generalized state spaces} defined by the allowed local sets of operators.

Generalized separability is useful if the operators in the generalized state spaces are constrained to possess some form of {\it generalized positivity}. We consider a number of different notions of generalized positivity in this work, the first of which is defined as follows. Suppose that on the constituent particles of a multiparty quantum state we may only make a restricted subset of local single particle measurements $ \mathcal{M}=\{M_i\}$, where each $M_i$ is a full POVM of measurement operators $M_i:=\{X^i_j|X^i_j\geq 0, \sum_j X^i_j = \I\}$. We define the \textit{dual} $ \mathcal{R} $ \cite{footnote1} to be the set of operators that give valid probabilities under Born's rule for these measurements,
\begin{equation}
\mathcal{R}=\{O \mid \ 0\le \mathrm{tr}(O X^i_j)\le 1,~\forall~X^i_j\}.
\end{equation}
We say that the elements of $\mathcal R$ are $ \mathcal{M} $-\textit{positive}. If an operator $O$ satisfies these inequalities strictly:
 \begin{equation}
 0 < \mathrm{tr}(O X^i_j) < 1,~\forall~X^i_j
\end{equation}
 then we say that $O$ is {\it strictly} $ \mathcal{M} $-positive, or {\it strictly} on the interior of $\mathcal R$, or that $\mathcal R$ strictly contains $O$. Versions of these definitions play a central role in the study of generalized probabilistic theories with an algebraic structure \cite{BAG}. The dual set  $\mathcal R $ may contain operators that are non-quantum (e.g. due to negative eigenvalues), and hence cannot physically be prepared. Nevertheless, with respect to the measurements in $ \mathcal{M} $, the elements of $ \mathcal{R} $ can serve a useful purpose. Suppose for example that a quantum entangled state $\Psi$ is known to have a separable decomposition in terms of $ \mathcal{M} $-positive operators,
\begin{equation*}
\Psi = \sum_n p_n A_n \otimes B_n \otimes C_n\otimes \cdots,  \,\,\, \mathrm{and} \,\,\, A_n,B_n,C_n,\cdots\in \mathcal{R},
\end{equation*}
where $p_n$ forms a probability distribution, then we say that $\Psi$ is $\mathcal R$-\textit{separable}. If this holds, then it implies that measurements from $\mathcal M$ will have a local hidden variable model. Moreover, if the probabilities $p_n$ can be efficiently sampled classically, then the outcomes of local measurements on $\Psi$ from $ \mathcal{M} $ will be efficiently simulatable (e.g. using the algorithm of \cite{HN03}).

This is the central strategy that we employ in this work. We consider arbitrary sets of local measurements $\mathcal M$ whose duals $\mathcal R$ strictly contain a pure quantum state. This allows us to employ continuity arguments to prove the existence of multiparty entangled states that are $\mathcal R$-separable. We will use the PEPS formalism to construct such pure entangled quantum states, and then modify the recipe to ensure that the weights $p_k$ can be efficiently sampled classically. We now describe the first of these constructions.

The PEPS formalism describes quantum states on a lattice in terms of underlying `{\it virtual}' (or `{\it valence}') bonds. Consider a lattice of $N$ sites of degree $v$ in which quantum systems of $d$-levels are positioned at each site $ s $, and lattice edges $ e $ signify neighbors (for simplicity we assume constant degree, but it is not essential to our arguments). In the PEPS formalism, a state of the system can be described by first allocating a pair of virtual $D$-level particles, in a maximally entangled state $\ket{\phi_D} =  \frac{1}{\sqrt{D}} \sum_{j}\ket{jj}$, to each edge on the lattice, with these virtual particles situated on neighboring sites.
The physical (but un-normalized) quantum state is described as the output of a local linear transformation $\mathcal{A}_s$ that `projects' (this term is used even though $\mathcal{A}_s$ may not be a projector) the $v$ virtual particles at each site $s$ down to the physical space, $\mathcal{A}_s:~(\mathbb{C}^{D})^{\otimes v}\mapsto \mathbb{C}^{d}$. The overall transformation on all particles is denoted $\mathcal{A}=\otimes_s \mathcal{A}_s$. The number of virtual levels $D$ is called the {\it bond dimension}. While any pure state can have a PEPS representation with high enough $D$, a family of states on increasing $N$ is called a PEPS family only if $D$ grows in a mild way. The cluster states are an example of a PEPS family with $D=2$ \cite{VC04}.

In order to reformulate PEPS from a generalized state space viewpoint, we need two types of generalized state spaces: one for the physical lattice, and others for the virtual lattice.
The set of operators $ \mathcal{R} $ will be the state space used for every particle in the physical lattice. For the virtual lattice, however, our strategy will be to pick a collection of state spaces $\mathcal{V} = \{\mathcal{V}_i \} $ (one for each virtual particle $i$) satisfying four useful properties: (i) they must be Hermitian, (ii) the operators in each set $\mathcal{V}_i$ must have {\it strictly} positive overlap with a pure state $\ket{\phi_i}$, i.e. $\bra{\phi_i}V\ket{\phi_i} >0 \, \forall \, V \in\mathcal{V}_i$, (iii) each virtual bond must be separable with respect to the corresponding pairs of these spaces, (iv) there must be a PEPS transformation $\mathcal{A}$ mapping to a pure quantum entangled state, such that the generalized separability of the virtual bonds leads to $ \mathcal{R} $-separability of the quantum entangled PEPS.

In order to find linear transformations that satisfy (iv) we need to know whether a given $\mathcal{A}$ leads to an $ \mathcal{R} $-separable state. A sufficient condition for this is that $\mathcal{A}$ is $(\mathcal{R},\mathcal{V})$-\textit{positive}, by which we mean that $\mathcal{A}$ takes products of inputs from the virtual state spaces at each site to operators $O$ such that $O/\tr(O) \in \mathcal{R}$ and $\tr(O) > 0$. If this holds for a given $\mathcal{A}$ then the PEPS will be $ \mathcal{R} $-separable, even though it could be quantum entangled. Note that $\mathcal{A}$ does not usually preserve normalization, and so the condition $\tr(O) > 0$ is used to ensure that the postselected probabilities (see e.g. Eq. (\ref{probdist})) are positive.

In order to find such suitable choices for $\mathcal{V}$ and $\mathcal{A}$, we first need to understand the kinds of state spaces for which the bond states $\ket{\phi_D}$ are separable.
 In \cite{AJRVprep} it was shown that given any set of $D^2$ orthogonal Hermitian operators $ \{C_1,\dots,C_{D^2}\} $ normalized to $ \mathrm{tr}(C_kC_l)=D\delta_{kl} $, the quantum state $ \ket{\phi_D}$ admits a separable decomposition as
\begin{equation}
\label{phi_sep}
\ket{\phi_D}\bra{\phi_D}=\frac{1}{D^2}\displaystyle\sum_{k=1}^{D^2}C_k\otimes C_k^T.
\end{equation}
Moreover, it was also shown that local state spaces which are strictly smaller than the convex hull of $\{ C_k \}$ cannot admit a separable decomposition of $\ket{\phi_D}$ \cite{foot6}.
The requirement (ii) above that each $\mathcal{V}_i$ have strictly positive overlap with a pure state can be imposed by simply picking a $\ket{\phi_i}$ and choosing the $C^i_k$s (where the upper index $i$ labels the virtual particle $i$) appropriately using the Gram-Schmidt process \cite{foot7}.

The fact that the transposition is required on one of the particles in Eq. (\ref{phi_sep}) means that the spaces at each end of a bond pair are different, and hence give each bond pair an orientation. While this will not affect our discussion significantly, it places a restriction that is important in constructing translationally invariant examples.

Using these state spaces for the virtual particles it is possible to write down a pure quantum entangled PEPS that is $\mathcal{R}$-separable given any choice of $\mathcal{R}$ that strictly contains a pure quantum state \cite{footnote2} as long as $d \geq 2^v$:

{\bf Recipe 1}: Let $\ket{\psi}$ be a pure $d$-level quantum state, with $d \geq 2^v$, that is strictly from the interior of the dual $\mathcal{R}$ of a set of local measurements $\mathcal M$. Pick the bond dimension of the virtual bonds to be such that $2 \leq D \leq d^{1/v}$. For each site $s$ define the rank-1 Kraus operator $Q^s=\ket{\psi}\bra{\alpha}$, where $\ket{\alpha}$ is a product of the $\ket{\phi_i}$ corresponding to the $\mathcal{V}_i$ at that site, i.e. satisfying (ii) above, $\bra{\phi_i}V\ket{\phi_i} >0 \, \forall \, V \in\mathcal{V}_i$. The Kraus operator $Q:=\otimes_s Q^s$
on the whole system has a corresponding completely positive (CP) map which is also {\it strictly} $(\mathcal{R},\mathcal{V})$-positive, as it takes product states from the $\mathcal{V}_i$ at each site to states from $\mathcal{R}$ with finite probability. Hence, by continuity we may pick a full rank Kraus operator $\wt{Q}^s$ for each site that is a small enough perturbation of $Q^s$, such that the corresponding CP map is also $(\mathcal{R},\mathcal{V})$-positive. As $\wt{Q}^s$ is a Kraus operator of full rank, the resulting PEPS will definitely be a pure entangled quantum state. However, as it is also $\mathcal{R}$-separable it will have a local hidden variable model with respect to measurements from $\mathcal{M}$. Note that this recipe extends to an arbitrarily large number of particles, and can be implemented in a translationally invariant way if the lattice bonds can be given a translationally invariant orientation (which is possible for a many even degree lattices). $\blacksquare$

Although this recipe gives a method for writing down a multi-particle PEPS that is $\mathcal{R}$-separable, and hence has a local hidden variable model for measurements from $\mathcal{M}$, it does not imply that outcomes from $\mathcal{M}$ can be efficiently classically sampled on these states. We now turn to this problem.

\textit{Efficient sampling:} As before, let us use local virtual state spaces $\mathcal V_i$ that are the convex hull of the $\{C^i_k\}$ operators from Eq. \eqref{phi_sep}. The virtual bonds are separable with respect to $\{C^i_k\}$ and their transpositions. For simplicity of notation we drop the transposition, and absorb it into the definition of the linear projection $\mathcal{A}$. We also omit the virtual particle label $i$.
In what follows, we use the fact that in the separable decomposition of $\ket{\phi_D}$ each product term contains the same operator for both particles. For each edge $e$ in the lattice, let $i_e \in 1,\dots,D^2$ be an {\it edge index} representing a choice of basis operator $C_{i_e}$. This allows us to write the virtual bond at edge $e$ as:
\begin{equation}
\ket{\phi_D}\bra{\phi_D} = {1 \over D^2} \sum_{i_e=1,\dots,D^2} C_{i_e} \otimes C_{i_e}.
\end{equation}
We use these edge indices to build up a description of the virtual bond state of the whole lattice. Each site is uniquely specified by the edges that are incident upon it, so to each site $s$ let us allocate a subset of edge indices, $\mathcal{I}_s:=\{i_{e_a},i_{e_b},\dots\}$, where $e_a,e_b,\dots$ are the incident edges at the site, and define an operator $C^{\mathcal{I}_s}:=C_{i_{e_a}}\otimes C_{i_{e_b}}\otimes \cdots$ on the virtual particles at site $s$. Then the whole virtual bond state is represented by:
\begin{equation}
\ket{\phi_D}\bra{\phi_D}^{\otimes E} = {1 \over D^{2 E}}\sum_{i_{1},\dots,i_E} \bigotimes_s C^{\mathcal{I}_s},
\end{equation}
where $E$ is the number of edges. Note that the operators on different sites can be related, as two sites $s,s'$ with a connecting edge must have a common edge index in $\mathcal{I}_s,\mathcal{I}_{s'}$.

The linear transformation $\mathcal{A}_s$ that acts at site $s$ projects the virtual particles down to a single state space. Let $O^{\mathcal{I}_s}:=\mathcal{A}_s(C^{\mathcal{I}_s})$, then the PEPS on the whole lattice can be written in the following way:
\begin{eqnarray}
\sum_{i_{1},\dots,i_E}  {\Pi_s \tr(O^{\mathcal{I}_s}) \over T~D^{2 E} } \bigotimes_s {O^{\mathcal{I}_s} \over \tr(O^{\mathcal{I}_s})}, \label{orep}
\end{eqnarray}
where $T$ is a an overall positive normalization factor that arises from the fact the overall transformation $\mathcal{A}$ acting on all sites does not preserve normalization. In this representation, the $(\mathcal{R},\mathcal{V})$-positivity of $\mathcal{A}$ is equivalent to:
\begin{eqnarray}
\tr(O^{\mathcal{I}_s}) > 0, \quad \mathrm{and} \quad {O^{\mathcal{I}_s}\over \tr(O^{\mathcal{I}_s})} \in \mathcal{R} \,\,\,\,\, \forall \, s. \nonumber
\end{eqnarray}
If these two conditions hold then Eq. \eqref{orep} will correspond to a state that is $ \mathcal{R}$-separable. We may easily write down linear operators that satisfy these conditions: simply pick operators $O^{\mathcal{I}_s}$ with positive trace that when normalized are inside $\mathcal{R}$. However, one would still need to check that the corresponding transformations $\mathcal{A}_s$ do indeed make a quantum state. One way of guaranteeing this is to also add a condition that the transformation $\mathcal{A}$ 
(with the transpositions momentarily taken out of its definition again) be a CP map, a condition which can in principle (although this may be difficult) be imposed by requiring the non-negativity of its Choi-Jamio\l{}kowski representation.
However, even if we have transformations $\mathcal{A}$ that satisfy these conditions, we still need to understand when Eq. (\ref{orep}) enables an efficient classical simulation. One way of using Eq. (\ref{orep}) to classically sample the outcomes of $\mathcal{M}$ measurements is to first sample the probability distribution over the edge indices
\begin{eqnarray}
p(i_{1},i_{2},i_{3},\dots,i_E):={\Pi_s \tr(O^{\mathcal{I}_s}) \over T \, D^{2 E} },\label{probdist}
\end{eqnarray}
and then sample outcomes of measurements from $\mathcal{M}$ on the corresponding operator product.
Unfortunately the first of these tasks is not straightforward. For each site $s$ the real numbers $\tr(O^{\mathcal{I}_s})$ can be viewed as coefficients of a classical tensor network with indices given by each edge index incident at the site $s$, and in general the resulting probability distribution (\ref{probdist}) cannot always be efficiently sampled (it can encode complicated partition functions, see e.g. \cite{EV14}).

However, some classical tensor networks can be efficiently sampled. Here we consider the simplest such case: if the tensors factorize as $\tr(O^{\mathcal{I}_s})=u_{i_{e_a}}w_{i_{e_b}}\dots$ for some positive vectors $\ve u, \ve w,\dots$, then the distribution can be efficiently sampled classically because it is a product of independent distributions over each edge. We now explain how to use this fact to construct classically efficiently simulatable examples.
We are looking for local linear transformations $\mathcal{A}_s$ at each site such that: (i) $\mathcal{A}_s$ is represented by one Kraus operator $\mathcal{A}_s(\cdot) = K_s^{\dagger} \, \cdot \, K_s$ (so that the PEPS is pure), (ii) each $K_s$ has full rank (so that the PEPS is guaranteed to be quantum entangled), (iii) each $\mathcal{A}_s$ is $(\mathcal{R},\mathcal{V})$-positive (so that the PEPS  is $\mathcal{R}$-separable), and (iv) the traces of the output operators factorize as $\tr(O^{\mathcal{I}_s})=u_{i_{e_a}}w_{i_{e_b}}\dots$ (so it can be efficiently sampled).
A method for constructing such transformations is as follows. It works for any $\mathcal{R}$ that strictly contains a pure state, as long as $d \geq 2^v$:

{\bf Recipe 2}: Let $\ket{\psi}$ be a $d$-level pure quantum state that is strictly from the interior of $\mathcal{R}$, and assume that $d \geq 2^v$. Pick the bond dimension to be $D=2$. Let $Q^s:=\ket{\psi}\bra{\alpha}$
be the the same operator used in Recipe 1, but instead write it as $Q^s=\ket{\psi}\bra{0,0,0,\dots}$ by appropriately choosing the local Hilbert space basis for each virtual particle. As with Recipe 1 the Kraus operator $Q:=\otimes_s Q^s$ gives a CP map that is {\it strictly} $(\mathcal{R},\mathcal{V})$-positive. Any small enough perturbation of $Q$ will maintain positivity, so we simply need to find a small perturbation of $Q$ that gives outputs with traces that factorize. Let $\ket{1}$ be an orthogonal state to $\ket{0}$ for each virtual particle, and introduce a $v$-component bit string $y$.
Now consider a Kraus operator for site $s$ of the form $\wt{Q}^s = \sum_y \epsilon^{\mathrm{Ham}(y)} \ket{\psi_y}\bra{y}$, where $\mathrm{Ham}(y)$ is the Hamming weight of $y$, and $\ket{\psi_y}$ is a set of $2^v$ orthonormal vectors in the Hilbert space of the $d\geq 2^v$ level physical particle at site $s$, which is picked such that $\ket{\psi}=\ket{\psi_0}$. For small $\epsilon$ this is close enough to $Q^s$, it is of full rank $2^v$ (and therefore makes an entangled quantum state), and the output traces can be readily verified to factorize \cite{footnote3}.
$\blacksquare$

\textit{How quantum entangled can these constructions be?}
Given any set of measurements whose dual strictly contains a pure state our recipes provide quantum entangled PEPS with local hidden variable models that can be efficiently simulated classically. However, it is important to understand {\it how} quantum entangled these states are. Given the lack of a simple framework for describing multipartite quantum entanglement, we will take an operational viewpoint and consider what tasks can be achieved by removing the restriction on local measurements.

Let us consider non-locality first. In \cite{PR92} it was shown that {\it any} pure multiparty entangled state is non-local, and hence the states constructed in Recipe 1 do not have a LHV model for all measurements. Moreover, this non-locality has some multipartite features: in general it cannot be allocated to some subset of the particles, because we are able to choose our states to be translationally invariant \cite{footnote4}.

However, it would be interesting to know if it is possible to construct families of $\mathcal{R}$-separable pure states for which measurements from $\mathcal{M}$ can be classically efficiently sampled, but if we allow all measurements then we can allow MBQC or infinite entanglement length \cite{Elength,Raussendorf}. We have not yet been able to answer this question for all $\mathcal{M}$. However, for specific choices of $\mathcal{M}$ it is possible to write down proof-of-principle examples that do this, and we will also outline a method that might enable progress more generally.

Consider for instance a PEPS with a trivial identity projector on virtual qubits. Each site is hence a quantum particle of $d=2^{v}$ levels, and fixes $D=2$. On this system we can efficiently sample local measurements $\mathcal M$ that are the dual of the $v$ virtual states spaces at each site. If we choose these virtual state spaces to consist of the unit trace phase point operators, as described in \cite{AJRVprep}, then $\mathcal M$  contains interesting classes of measurements that are not local at the level of the virtual particles at a specific site (including for example Bell measurements on any two virtual particles at the same site, measurements that are multiparty entangled over the $d$ virtual qubits at a single site, and non-stabilizer measurements). As far as we are aware no previously proposed classical simulation technique can sample them efficiently. However, the system still contains a great deal of quantum entanglement. Through entanglement swapping using Bell basis measurements on the virtual particles at the same site we can see that the system has an infinite entanglement length (even though Bell measurements are in $\mathcal M$), and for most lattices the state is universal for quantum computation using appropriate local measurements outside $\mathcal M$. This example shows that generalized-separability can enable classically efficient sampling of systems that can still contain enough quantum entanglement to perform MBQC with unrestricted measurements.

A generalization of this argument could probably be made to work for other classes of measurement with $d \geq 2^v$. Suppose that we have a set of measurements $\mathcal M$ on a $d \geq 2^v$ level physical particle, and that $\ket{\psi}$ is a pure state strictly on the interior of $\mathcal R$. Pick the virtual state spaces and their computational basis such that $\ket{\psi}=\ket{0,0,0,\dots}$. If we set the virtual bonds in a state $\sqrt{1-\delta^2}\ket{00}+\delta\ket{11}$ where $\delta$ is small enough so that the virtual bonds form an $\mathcal R$-separable state, and if we take $\mathcal A$ to be the identity, then the resulting lattice will be efficiently samplable for $\mathcal M$ measurements. However, if the lattice degree $v$ is big enough, then using procrustean entanglement distillation \cite{BBPS96} with percolation arguments (see e.g. \cite{Eperc}) would probably enable distillation of a lattice of Bell state bonds that is connected enough to do MBQC. The difficult part of making these arguments work would be to understand the interplay between how far on the interior $\ket{\psi}$ is, and the lattice degree required, but it seems likely that it would work for a wide variety of $\mathcal M$.

\textit{Conclusions:}
Given any set of local quantum measurements satisfying the restriction that their dual strictly contains a quantum pure state, we have described families of pure entangled state with local hidden variable models and classically efficient simulations. Of course this restriction cannot be removed in our approach: if the dual does not strictly contain a pure state, it has to be the set of quantum states (it cannot be smaller as the quantum states are in every POVM's dual), and separability reduces to the usual quantum version. This also suggests that the continuity argument is essential for the existence proof, as the constructions work for measurement sets that are arbitrarily close to the full quantum set. There is scope, however, for expanding upon Recipe 2 by considering alternative classical tensor networks with efficient classical sampling algorithms for defining the probabilities (\ref{probdist}). Moreover, it seems likely that modified constructions could remove the requirement that $d \geq 2^v$.

Although we have provided existence arguments that can be applied to many choices of $\mathcal{M}$, it will be interesting to apply these ideas to concrete situations in which we have a specific choice of $\mathcal{M}$ or $\mathcal{A}$ in mind.

If, for example, we have a specific choice of $\mathcal{M}$, then our goal would be to identify choices of $\mathcal{V}$ for which the largest sets of $ \mathcal{A} $ would be $(\mathcal{R},\mathcal{V})$-positive. Although one way of doing this could be using continuity approach (if a quantum state $\ket{\psi}$ is deep enough in the interior of $\mathcal{R}$, then the further away from rank-1 the perturbations $\wt{Q}^s$ can be), one could also attempt a direct search for $(\mathcal{R},\mathcal{V})$-positive PEPS transformations $ \mathcal{A} $.

On the other hand we could consider situations in which $\mathcal{A}$ is fixed, but instead seek choices for $\mathcal{V}$ that maximize the set of local measurements $\mathcal{M}$ for which $\mathcal{A}$ is $(\mathcal{R},\mathcal{V})$-positive. Examples of important $\mathcal{A}$ include those that make AKLT \cite{AKLT} states and cluster states. When defined on appropriate lattices, these states are universal for quantum computation, so it is unlikely that some choices of $\mathcal{V}$ can be found that allow $\mathcal{M}$ to include pure quantum measurements in more than a few directions. However, it will be interesting to see what noisy measurements are allowed (e.g. in order to understand bounds on fault-tolerant quantum computation), and how $\mathcal{M}$ can change for variants of PEPS that can be used to describe interesting many-body phenomena such as phase transitions.

\bigskip

\section{Acknowledgments}

We thank Miguel Navascues and Terry Rudolph for valuable discussions, and Marco Piani for valuable comments and pointing out an error in an earlier draft of this work. This work was supported by EPSRC grant EP/K022512/1.


\begin{thebibliography}{99}
\bibitem{footnote1} We use this not standard notation because $\mathcal{R}$ signifies `real' as opposed to `virtual' particles.
\bibitem{foot6} Note that it is in our interest to use the smallest state spaces that we can (while still requiring separability of the virtual bonds), as if $ \mathcal{A} $ is $(\mathcal{R},\mathcal{V})$-positive for a given $ \mathcal{V} $, then it is also $(\mathcal{R},\mathcal{W})$-positive for $\mathcal W= \{\mathcal W_i\}$ such that ${\mathcal{W}}_i \subseteq {\mathcal{V}}_i$.
\bibitem{footnote2} A similar requirement was used in a recipe for constructing entangled states with local hidden variable models in \cite{RV10}, however, in contrast to the constructions of \cite{RV10} our states do not tend to product states as $N$ is increased.
\bibitem{footnote3} This can be seen as follows. Consider an operator $A \otimes B \otimes\cdots$ transformed by $\wt{Q}^s$. The output trace is:
\begin{eqnarray}
\tr(\wt{Q}^s (A \otimes B \otimes \cdots) (\wt{Q}^s)^{\dagger}) = \sum_{y} \epsilon^{2\mathrm{Ham}(y)} \bra{y} A \otimes B \otimes\cdots \ket{y} \nonumber \\
= (\bra{0} A \ket{0} +\epsilon^2 \bra{1} A \ket{1} )(\bra{0} B \ket{0} +\epsilon^2 \bra{1} B \ket{1} )\cdots \nonumber
\end{eqnarray}

which is clearly factorized.
\bibitem{footnote4} In this context we speculate that it will be interesting to consider lattices that are expander graphs with fixed degree: if we accept the intuition that such graphs are an approximation to a fully connected graph, then for such systems the non-locality could arguably be a highly `multipartite' property.
\bibitem{foot7} This can be seen as follows. Assume that for virtual particle $i$ we pick $\ket{\phi_i}=\ket{0}$. We can satisfy the requirement by picking $k=1,\dots,D^2$ operators satisfying $\bra{0}C^i_k\ket{0}=\sqrt{D}c$ for some $c>0$. Using Gram-Schmidt, we can construct any $D^2 \times D^2$ orthogonal matrix with all elements in the top row given by $c$, where $c$ is fixed by the row normalisation of the orthogonal matrix. Use column $k$ of this matrix as the expansion of $C^i_k/\sqrt{D}$ in any orthonormal basis of operators containing $\ket{0}\bra{0}$.
\end{thebibliography}
\end{document}